**RESEARCH**

# Visualization of Large Volumetric Multi-Channel Microscopy Data Streams on Standard PCs

Tobias Brix[*], Jörg-Stefan Praßni and Klaus Hinrichs

**Abstract**

**Background:** Visualization of multi-channel microscopy data plays a vital role in biological research. With the ever-increasing resolution of modern microscopes the data set size of the scanned specimen grows steadily. On commodity hardware this size easily exceeds the available main memory and the even more limited GPU memory. Common volume rendering techniques require the entire data set to be present in the GPU memory. Existing out-of-core rendering approaches for large volume data sets either are limited to single-channel volumes, or require a computer cluster, or have long preprocessing times.

**Results:** We introduce a ray-casting technique for rendering large volumetric multi-channel microscopy data streams on commodity hardware. The volumetric data is managed at different levels of detail by an octree structure. In contrast to previous octree-based techniques, the octree is built incrementally and therefore supports streamed microscopy data as well as data set sizes exceeding the available main memory. Furthermore, our approach allows the user to interact with the partially rendered data set at all stages of the octree construction. After a detailed description of our method, we present performance results for different multi-channel data sets with a size of up to 24 GB on a standard desktop PC.

**Conclusions:** Our rendering technique allows biologists to visualize their scanned specimen on their standard desktop computers without high-end hardware requirements. Furthermore, the user can interact with the data set during the initial loading to explore the already loaded parts, change rendering parameters like color maps or adjust clipping planes. Thus, the time of biologists being idle is reduced. Also, streamed data can be visualized to detect and stop flawed scans early during the scan process.

## Background

Visualization of multi-channel microscopy data plays a vital role in biological research. With the ever-increasing resolution of modern microscopes the data set size of the scanned specimen grows steadily. On commodity hardware this size easily exceeds the available main memory and the even more limited GPU memory. Since common volume rendering techniques require the entire volume to be present in the GPU memory, out-of-core rendering approaches have been introduced. Although nowadays petabyte-sized data sets can be visualized [1], existing out-of-core rendering approaches are either limited to single-channel volumes [2], or require a computer cluster [1], or have long preprocessing times [3]. Throughout this paper we will use the term *RAM* for the CPU memory and *GPU memory* for the graphic board's main memory.

In the field of biological research specimens are scanned by microscopes to create a digital volumetric representation. In the visual analysis of these specimens usually multiple scanned volumes have to be rendered into a single image. For instance when using a fluorescence microscope, the specimen is illuminated with light of specific wavelengths which are absorbed by the fluorophores, causing them to emit light of longer wavelengths. In this manner, the distribution of a single fluorophore is imaged at a time. Multi-channel volumes of several types of fluorophores must be composed by combining several single-channel volumes. Important biological research results are often based on the visual study of these multi-channel volumes, especially in cell biology.

The data sets we are using to evaluate our technique (see Sec. *Results and Discussion*) are mouse embryos scanned by a Single Plane Illumination Microscope or Selective Plane Illumination Microscope (SPIM) ([4],[5]), which are special kinds of fluorescence microscopes. Each data set contains three channels of different light waves. Each channel has the same resolution and orientation which is typical for multi-channel microscopy volumes. Microscopes scanning the specimen slice-wise, like SPIM, can be used

---

[*]Correspondence: t.brix@wwu.de
University of Münster, Einsteinstr. 62, 48149, Münster, Germany
Full list of author information is available at the end of the article for streaming the scanned data. After a slice has been scanned, it can be submitted to the proposed system that immediately updates the rendering. So the scientist is able to visually analyze the scanned data set during the often lengthy scan process and may interrupt it, if a misconfiguration is detected.

Discussions with biologists about their analysis workflow of multi-channel microscopy data have revealed two main issues, which they wanted to be addressed. The first are the demanding hardware requirements. The commercial software they use needs to load the entire volume into the RAM. Therefore, they have to share expensive high-end workstations with 64 GB or more RAM only for the purpose of visual analysis and image creation. In their opinion the visualization of the volumetric data should be possible on their standard PCs with normally up to 16 GB RAM. The other issue are the long preprocessing times. The time of loading the different channels into the software until the first image has been rendered usually causes a workflow interruption of up to 10 minutes.

Taking into account these issues, we defined the following design goals for our rendering technique:

1. Out-of-core visualization of multi-channel volumetric microscopy data, including support for data sets exceeding the RAM.
2. Interactive preprocessing which allows the user to interact with the data while it is being loaded.
3. Support for visualizing streamed data in order to detect flawed scans early on.
4. No high-end system requirements.

There are some existing software tools which can be used by biologists to visualize their microscopy data. We just want to point out some of them. For instance the open source *ImageVis3d/Tuvok* system [6] developed at the University of Utah which is based on distributed rendering [7]. This tool is designed for biomedical visualization and supports most common analysis tools, e.g, clipping planes. However, the software needs to convert the volume file into the *universal volume file* (UVF) which results in violations of our design goals of supporting streamed data without preprocessing.

Another free available tool also developed at the University of Utah is *FluoRender* [8]. The software was designed for visualizing multi-channel microscopy data optimized for the workflow of neurobiologists [9]. It supports segmentation tools [10] and can integrate simple polygon geometries into the rendering, e.g., as crude region definition. However, FluoRender can not visualize streamed data and can not visualize data sets exceeding the RAM or GPU Memory.

A commercial software for visualizing microscopy data is *Volocity* by *PerkinElmer Inc (Massachusetts, USA)* [11]. The software is capable of rendering multi-channel data sets whose size exceeds the GPU Memory. Many analytic tasks are supported, e.g., measurement of distances inside the data set or tracking of objects during time steps. However, the visualization of streamed data is not supported.

Another popular commercial software used by biologists we want to mention is *Imaris* by *Bitplane AG (Zurich, Switzerland)* [3]. Imaris is also capable of rendering GPU Memory exceeding data sets and even the RAM exceeding ones. However, volume files have to be converted into the native Imaris file format, which again results in preprocessing and again no support of streamed data.

In order to visualize large volume data sets exceeding the GPU memory, out-of-core and multi-resolution approaches have been proposed. Hierarchical approaches have been presented by Boada et al. [12] and Guthe et al. [13], who employ a wavelet representation and use screen-space error estimation for level of detail selection. We do not follow the approach of volume compression, since usually these methods result in expensive preprocessing (goal 2) or are not feasible for dynamic updates (goal 3). Probably we will consider these approaches in future work, see Sec. *Conclusion*. Instead, we based our technique on a spatial hierarchical octree bricking scheme. We just address the most important work related to our approach and refer to the survey about octree based volume rendering by Knoll [14] for a more detailed discussion. Among the first developers of this kind of octree structure were LaMar et al. [15] and Weiler et al.[16]. Two of the more popular implementations using this approach were introduced by Gobbetti et al. [17] and Crassin et al. [2]. The Crassin implementation was called *GIGAVOXEL* and has been redesigned and optimized in the *CERA-TVR* framework by Engel [18].

We share the basic ideas these papers are based on, but modified and improved them to solve our special biological task. We want to point out that all mentioned previous systems only support single-channel volume rendering and need the entire volume data set to be present during the construction of the octree. Thus their techniques can not be used to visualize streamed data.

Systems which support data streams were introduced by Hadwiger et al. [1] and Beyer et al. [19]. These systems are capable of rendering microscopy data with a size up in the petabyte range. Such huge data sets are generated for instance in neuroscience by electron microscopes [20]. Therefore, the approaches of Hadwiger et al. and Beyer et al. use a network consisting of multiple high-end computers to achieve the needed computing power. This is against our premise to visualize the data on commodity hardware. However, such data set sizes are quite unusual in other fields of biological research like cell biology. Usually the data set size for one channel is less than 10 GB.
In the next section *Methods* we describe our visualization technique, which achieves the previously defined four goals.

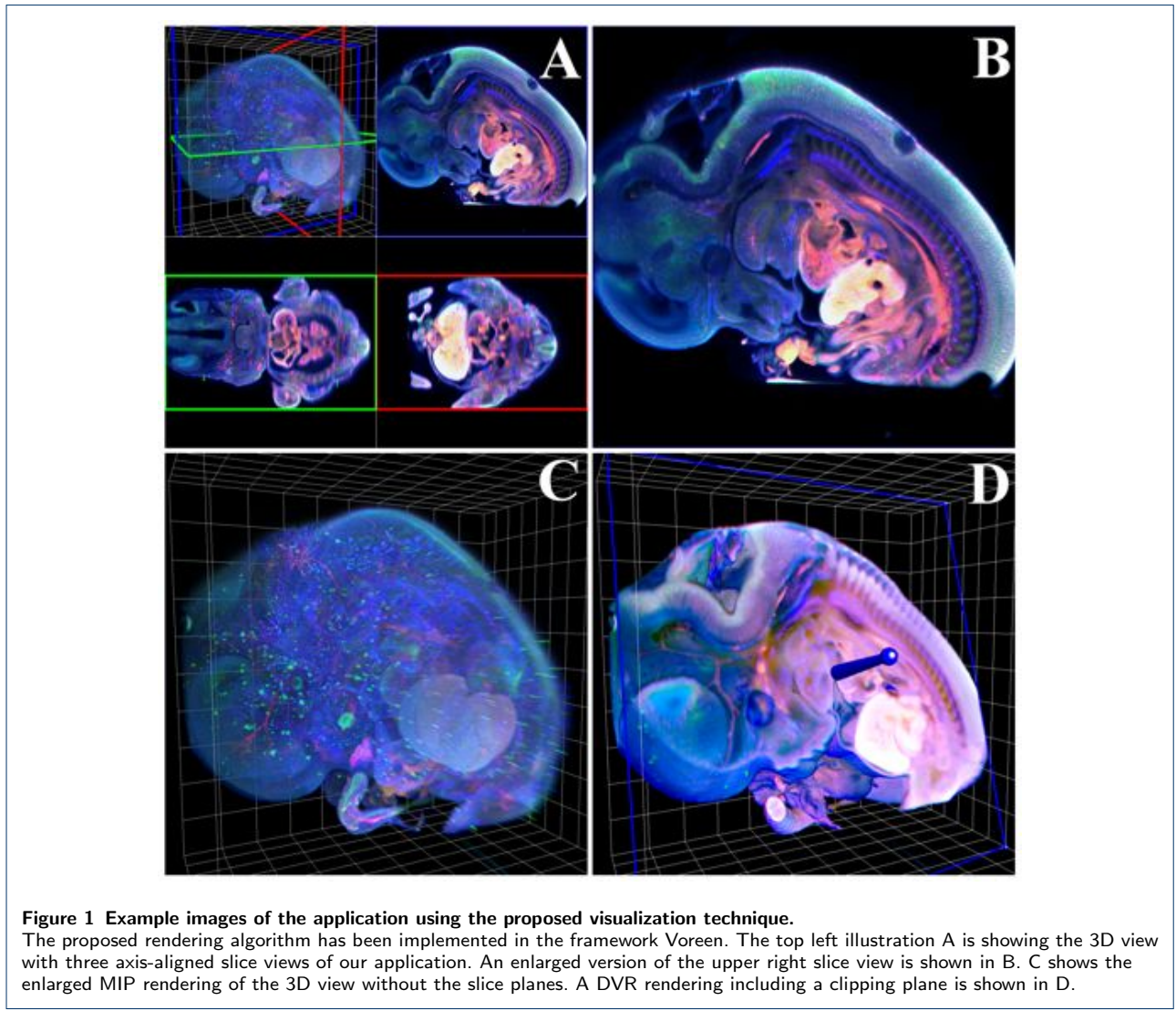

**Figure 1 Example images of the application using the proposed visualization technique.**
The proposed rendering algorithm has been implemented in the framework Voreen. The top left illustration A is showing the 3D view with three axis-aligned slice views of our application. An enlarged version of the upper right slice view is shown in B. C shows the enlarged MIP rendering of the 3D view without the slice planes. A DVR rendering including a clipping plane is shown in D.

*Integration into Voreen*
The proposed approach has been implemented in the framework Voreen (Volume Rendering Engine) [21]. Using this framework as backbone of our implementation enables the biologists to use all tools provided by Voreen. Example images of our system featuring the proposed technique can be seen in Fig. 1. The system consists of four linked views (Fig. 1 A). Each view, a 3D rendering and three axis-aligned slice views, can be enlarged separately. Colored planes in the 3D view indicate the position of the associated slices sharing the same colored border in the 2D views. Beside customizable color maps for each channel, arbitrary clipping with up to three clipping planes is supported (Fig. 1 D). The user can change the 3D rendering between direct volume rendering (DVR) and the maximum intensity projection (MIP), which is more common in biological research. Segmentations of the data set are supported by the integrated random walker algorithm by Praßni et al. [22]. Also prodution of videos is supported by the framework, including camera animations and any modification of rendering parameters, which is often used by our cooperation partners. The software is capable of loading several common volume formats, e.g., OME-TIFF or DICOM.

A released version of this application, not supporting the incremental construction yet, can be downloaded from the website `voreen.uni-muenster.de`.

## Methods

Our technique consists of three main components: the data structure, the GPU rendering using OpenCL, and the data streaming between different levels of the memory hierarchy.

*Data Structure:* We employ a multi-resolution hierarchy for handling volume data similar to the one suggested by Crassin et al. [2]. It consists of a 3D mipmap storing the voxel data at different levels of detail (LOD) as well as an octree providing access to this mipmap (see Fig. 2). The mipmap is organized in *bricks* of constant resolution (e.g., $16^3$ voxels). In the remainder of this paper we will call the entirety of these bricks the *brick pool*. Each octree node represents a subset of the volume space at a specific LOD and is associated with the corresponding brick. In contrast to Crassin et al. our approach supports multi-channel volumes and incremental construction, which allows us to handle data streams and enables the user to interact with the rendering during creation time.

*Rendering:* We use a GPU-based ray-casting approach, which has been extended for out-of-core rendering. While the entire octree can be kept in a memory efficient format on the GPU, usually only a subset of the brick pool can be stored in the GPU memory. Thus, the rendering algorithm has to request bricks on demand at the appropriate LOD and has to handle missing data appropriately.

*Data Transfer:* In general the brick pool size even exceeds the CPU RAM size and therefore has to be stored on mass storage like SSD or hard drive. In the remainder of this paper we will use equivalently the terms mass storage and disk. Bricks requested by the rendering algorithm have to be uploaded onto the GPU, with the CPU RAM serving as intermediate cache. Sine mass storage accesses are time-consuming, we use a paging approach to minimize disk interaction as shown in Fig. 5.

### CPU Data Structure

To simplify the description of the data structure, we assume a single-channel volume with a cubic resolution that is at least $N$ times evenly divisible by two, i.e.:

$$M \cdot 2^N \text{ with } M, N \in \mathbb{N}, M > 0, N \geq 0 \qquad (1)$$

This assumption can be dropped, as will be shown in Sec. *Handling Volumes of Arbitrary Dimensions*.
As illustrated in Fig. 2, the volume data is stored in a 3D mipmap containing $N+1$ levels with level zero denoting the original resolution. Each level is constructed by iterative down-sampling of the previous level. The mipmap is organized in *bricks* of constant resolution $M^3$. The entirety of all bricks is called the *brick pool*.

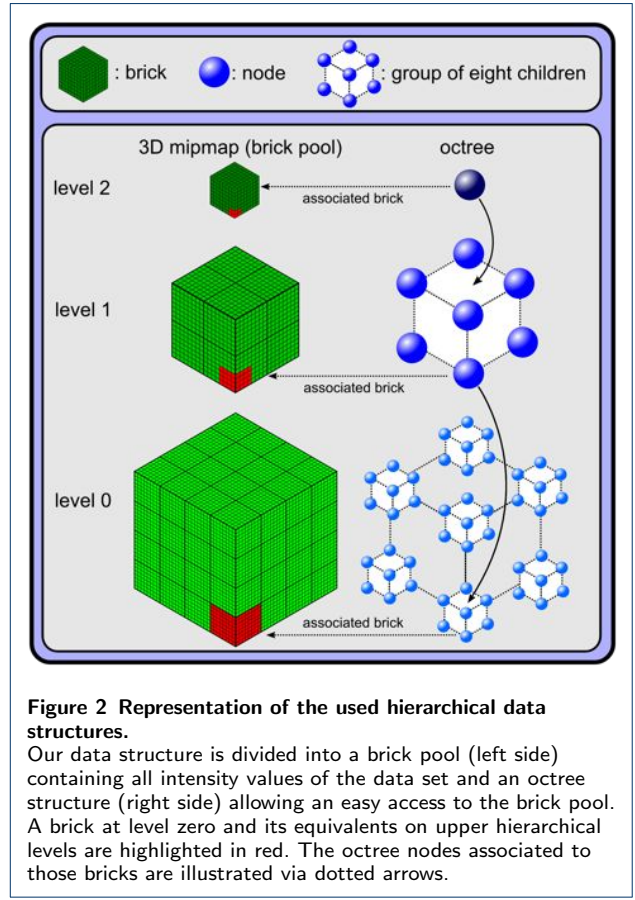

**Figure 2 Representation of the used hierarchical data structures.**
Our data structure is divided into a brick pool (left side) containing all intensity values of the data set and an octree structure (right side) allowing an easy access to the brick pool. A brick at level zero and its equivalents on upper hierarchical levels are highlighted in red. The octree nodes associated to those bricks are illustrated via dotted arrows.

In Fig. 2 one brick at level zero and its associated entries at the lower mipmap levels are highlighted in red. The memory required for storing the brick pool is only about 14% larger than the original volume data size, since each new level needs $\frac{1}{8}$th memory of the previous level. This results in a geometric series:

$$\sum_{i=0}^{N} \left(\frac{1}{8}\right)^i \xrightarrow{N \to \infty} \frac{8}{7} \approx 1.14$$

To access the bricks of the brick pool we use an octree whose structure correlates to the mipmap structure: Each node of the tree stores a pointer to a brick. The root node references the brick of the highest mipmap level, which approximates the entire volume. Each of the eight child nodes represents an octant of the parent node volume space and points to a brick approximating the corresponding octant. Each leaf node of this $N+1$ level deep tree stores a pointer to the brick containing the volume data in the original resolution. The right side of Fig. 2 shows the octree, emphasizing the node path to the highlighted brick in the brick pool. The data structure can be optimized by specifying a *homogeneity threshold* to handle homogeneous regions, in particular empty space, inside the data set. We flag a brick as homogeneous, if the difference between the minimum and the maximum intensity value stored in the brick is below the homogeneity threshold. Such a homogeneous brick can be deleted from the brick pool to reduce the needed memory and can be approximated by its average intensity value (AVG value). In this case, the corresponding octree node does not store a pointer to a brick but only the AVG value. Furthermore, a subtree is homogeneous, if the difference between the minimum and the maximum intensity value stored in the entire subtree is below the homogeneity threshold. Such a subtree can be deleted from the octree structure to reduce the tree depth for homogeneous regions. A homogeneous subtree always implies the homogeneity of all contained bricks. These optimizations are not illustrated in Fig. 2, but the bricks flagged as homogeneous will be used in the next section.

*Incremental Construction*

The iterative down-sampling to construct the 3D mipmap requires the presence of the entire data set. In the application case of data streams emitted by microscopes only small parts of a data set are present at a time. To handle these data streams we propose an incremental octree construction. At any point of the construction we can insert a cuboid block of voxel data with an arbitrary resolution at an arbitrary position into the octree. Already inserted data can be overwritten. Previous knowledge about the data set is not required, only its channel count and final resolution need to be known in advance. However, this information should be present for already scanned data as well as for streamed data. In the case of slice-wise scanned microscopy data, the inserted blocks usually correspond to the scanned slices.

In the initialization step the root node of the octree storing only a background value is created. Thus, the root node is present from the beginning and the tree can be traversed at any time. The background value represents not yet inserted parts of the volume. The iterative block insertion consists of three steps:

1. Copy the values of the block to be inserted into the corresponding bricks associated with the leaf nodes of the octree. New needed nodes and bricks on the insertion path are allocated, and yet unknown brick values are set to the background value. It is a design choice to create all children of a node although only one child is needed. Thus, we have an easier handling of special cases, since each node can only have zero or eight children. Those not yet needed children simply store the background value and have no brick associated.
2. Update the bricks on the insertion path from the leaf nodes to the root. Each parent brick will be updated by half-sampling its eight child node bricks.
3. Delete homogeneous bricks on the insertion path and store their AVG value instead. If an entire subtree is homogeneous, remove all nodes of the subtree.

Fig. 3 shows an example of the incremental construction of a one-dimensional single-channel data set with a brick resolution of four and a homogeneity threshold of one. After the initialization the first block with values [1,5,2] is inserted. All needed nodes and bricks on the insertion path are created. The original intensity values are inserted at the leaf level, and all bricks on the path are updated. All updated values are highlighted in red. The already inserted data do not have to be coherent, as the second insertion of [3,3,2,2] shows. The first optimization takes place after the third insertion of [4,3,3]. Two bricks are flagged as homogeneous, since the difference of their minimal value (3) and their maximal value (3) is below the homogeneity threshold. Thus, they are deleted and only their AVG value (3) is stored. However, the left subtree can not be deleted, since the leftmost leaf node's difference between its minimum (1) and its maximum (5) value is not below the threshold. After the fourth insertion the entire volume data has been filled into the data structure. The root node is flagged as homogeneous, but no subtrees can be deleted.

We would like to point out that during an insertion each brick on the insertion path from the leaf nodes to the root must be updated which is the most time-critical part. Therefore, it has a positive performance impact to insert large, brick aligned blocks to reduce the number of parent brick updates.

*Handling Volumes of Arbitrary Dimensions*

Equation 1 in Sec. *CPU Data Structure* is of course a major limitation. Given a fixed brick resolution $M$, we virtually increase the volume resolution to satisfy this equation. For example, a volume with original resolution $3865 \times 1966 \times 2893$ would be virtually increased to a $4096^3$ volume for a given $M = 32$. The extra memory required by the virtual extension of the volume is minimal. No extra bricks have to be allocated, since the nodes outside the original volume border are homogeneous and just store the background value. Bricks which are partly outside the volume border need special treatment during the incremental construction. The brick values outside the original volume are not used to determine the minimal, maximal and average value. Also they are not used to be half-sampled into the parent brick.

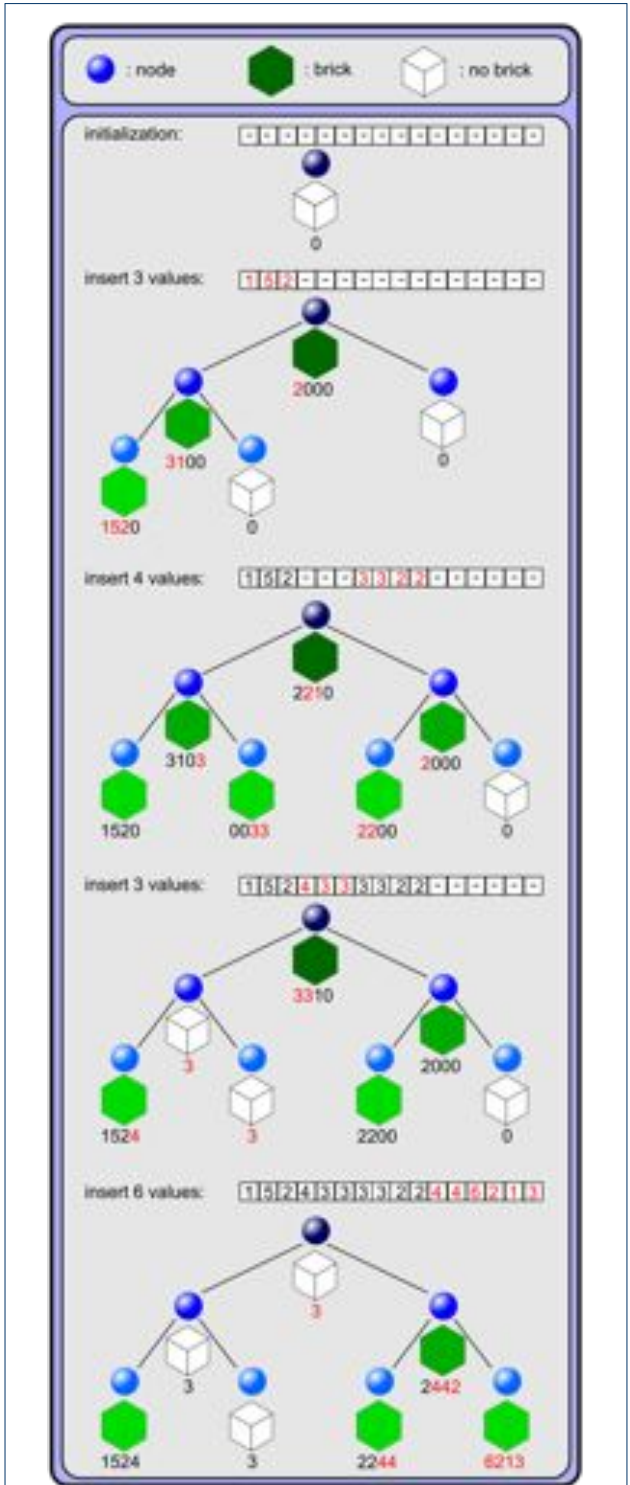

**Figure 3 Example of the incremental octree construction.** The illustration explains the incremental octree construction for a one-dimensional single-channel data set with a brick resolution of four and a homogeneity threshold of one. After initialization with an empty brick pool and only the root node, four blocks of intensity values are incrementally added. Each step highlights the added intensity values and the influence on the octree structure.

In practical use, we choose an even brick resolution $M$ between 16 and 64, usually a power of two although not necessarily, so that each calculation of a parent brick value only requires one child brick. For an odd resolution, up to all eight child bricks are required to calculate one parent brick value.

Furthermore, $M$ can be chosen individually for each dimension to prevent empty subtrees at the root level. In the previous example a brick resolution of $32 \times 16 \times 32$ results in a virtual volume resolution of $4096 \times 2048 \times 4096$ and removes an unnecessary split in the second dimension.

*Handling Multi-Channel Data*
The straightforward approach for handling multi-channel data would be to create a data structure for each channel. This approach results in two drawbacks: First, these multiple octrees would have to be traversed separately during ray-casting (see Sec. *Basic Ray-Casting*), which has a major impact on the rendering performance. Second, the multiple trees increase the GPU memory consumption (see Sec. *GPU Data Structure*).

In our special application case we can store all channels in the same data structure. Since all channels are sampled on a common grid, the octree does not have to be modified. Only the bricks have to be adapted. A brick has to contain all voxel information of all channels. This means a brick with resolution $M^3$ will have the size $M^3 \cdot C$ in the case of $C$ channels. The voxel data of the channels is stored interleaved inside the brick. This order creates a better memory coherence, which is preferable on parallel computing frameworks like OpenCL.

The incremental octree construction has to be modified to be capable of handling multi-channel data. When the first block of a channel is added to a newly allocated brick, the intensity values of the other channels are not yet known. Therefore, all other values in the brick are initialized with the background value. If a brick is homogeneous in each channel, we can delete it, but have to store the AVG value of each channel instead.

A disadvantage of this implementation is that a brick can be deleted only if all channels are homogeneous in that sub-region of the volume. However, we tolerate the potentially increased number of bricks, since we need to traverse only a single octree instead of one for each channel. Furthermore, the channels of the microscopy data sets used by our cooperation partners from biology are often quite similar in their distribution of empty space.

OpenCL Rendering
In this section we describe our rendering algorithm and the required data structure on the GPU.

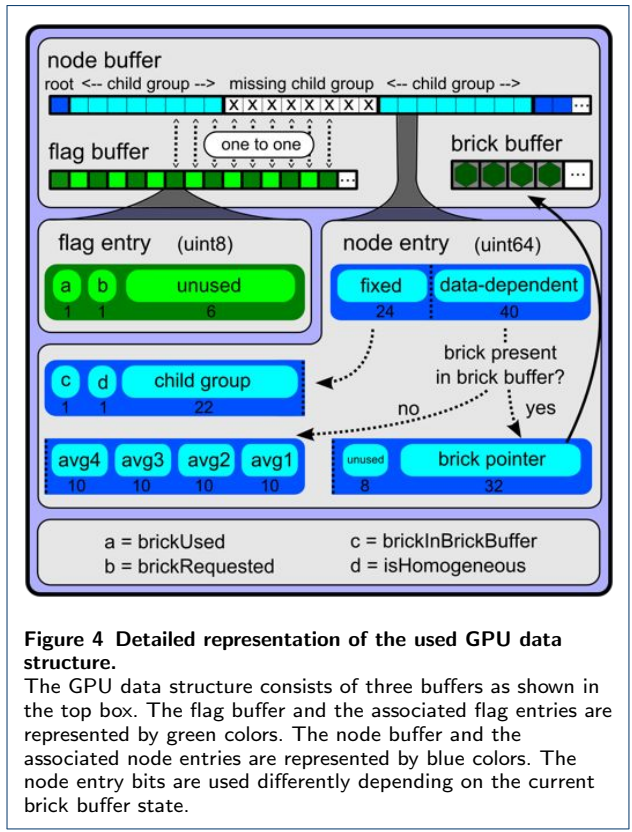

**Figure 4 Detailed representation of the used GPU data structure.**
The GPU data structure consists of three buffers as shown in the top box. The flag buffer and the associated flag entries are represented by green colors. The node buffer and the associated node entries are represented by blue colors. The node entry bits are used differently depending on the current brick buffer state.

*GPU Data Structure*
The rendering algorithm on the GPU needs three OpenCL buffers. The first two buffers, *brick buffer* and *node buffer*, are a mapping of the CPU data structure and are only read by the GPU. The third *flag buffer* is written by the GPU and is used as a feedback buffer to communicate with the CPU.

*Brick Buffer:* This buffer stores bricks of the brick pool, which are needed for the next rendering pass. A buffer size equal to the entire brick pool size would be preferable, but on commodity hardware we are limited to a buffer size between 256 MB and 4 GB.

*Node Buffer:* This buffer contains the CPU octree in a memory efficient format. We represent an octree node by a 64 bit integer value, a so called *node entry*, and we store the octree nodes in breadth-first order, i.e., beginning with the root node, all potential eight children of the node are stored in sequence. If a node has no children, the memory will be allocated anyway. Thus, children created later on by the incremental construction can be added without changing or shifting the other node entries. The node buffer for a tree with eight levels requires only 20 MB memory.

The 64 bits of a node entry are subdivided into an always used 24 bit part and a 40 bit part depending on the number of channels and rendering stats as shown in Fig. 4.

- The first bit of the always used first 24 bits indicates whether the brick associated to the octree node is present in the brick buffer.
- The second bit indicates whether the node is not homogeneous, i.e., the node has an associated brick.
- The next 22 bits are used as a pointer to the group of eight children inside the node buffer. The pointer is zero in case the node has no children. Since we need just a single pointer for a group of eight children, these 22 bits are enough to locate $8 \cdot 2^{22}$ nodes. Since a complete tree of $N$ levels contains $\sum_{i=0}^{N-1} 8^i$ nodes, we can address a tree of at most nine levels, e.g., assuming a brick resolution of 64 we can address a $16384^3$ volume with a memory consumption of approximately 8 TB.
- If the associated brick is present in the brick buffer, i.e., the first bit flag is set, the last 32 bits of the remaining 40 bits are used as a pointer to the brick in the brick buffer. Assuming each node of a nine level deep octree has an associated brick, 25 bits would be enough to address each brick. However, we allocate 32 bits to be byte-aligned, since the other 8 bits are not used in this case anyway.
- If a node is homogeneous or the brick is not present in the brick buffer yet, the 40 remaining bits are used to store the AVG values of each channel of this node. The 40 bits are divided equally among the channels. If the data set format requires more bits, e.g., 32 bit integer with two channels, a conversion by a bit length reduction has to be performed. However, since microscopy data is usually in uint16 format, this compression has no visible influence on the rendered image. Our approach can be extended to more than four channels, but would reduce the available number of bits per channel. Nevertheless, an extension to 128 bits per node entry would be possible at the cost of additional memory. This would also enable the storage of AVG values in other formats, e.g. float, without conversion.

*Flag Buffer:* The flag buffer consists of byte-sized *flag entries*, which are in a one-to-one correspondence to the node entries in the node buffer. The first two bits of each entry are used to encode whether the brick of the corresponding node entry has been used during the last rendering pass and whether it has been requested for the next rendering pass.

*Basic Ray-Casting*
The visualization itself is performed in our approach by GPU-based volume ray-casting introduced by Krüger et al. [23]. We use an out-of-core variant of their

**Algorithm 1:** Pseudo code of DVR ray traversal.

```
1  float4 resultCol = vec4(0.f);
2  float pos = entryPoint;
3  while pos < exitPoint do
4     float intensity;
5     NodeEntry node = getNodeAtPosition(pos);
6     for int c=0; c < Channels; c++ do
7        if node.isHomogeneous() then
8           intensity = node.getAVG(c);
9        else
10          if node.hasBrickInBuffer() then
11             setBrickUsed(node);
12             intensity = node.brick.valueAt(pos,c);
13          else
14             setBrickRequested(node);
15             intensity = node.getAVG(c);
16       resultCol = composite(resultCol,
                                applyTF(intensity));
17       pos += samplingStep;
```

approach as shown in Algorithm 1. Its main difference to standard ray-casting is the requesting and handling of missing subsets of volume data on the GPU. The following description refers to direct volume rendering (DVR). Nevertheless, the tree traversal also applies to other compositing modes such as maximum intensity projection (MIP).

At each step of the ray traversal the octree node containing the current sampling position at the *optimal LOD* has to be determined. Depending on the camera position the optimal LOD is calculated, such that the brick voxel size of the requested node is projected to the pixel size of the resulting image. Beginning at the root node, we iteratively descend to the child node whose bounding box contains the sampling position until the desired node level is reached. Since the volume space is divided equally at each tree level, a node's bounding box can be derived from its parent's bounding box. Thus, only the bounding box information of the root node, i.e., the entire volume, has to be stored while the bounding boxes of the remaining nodes can be calculated during descent. In case of an incomplete tree, e.g., during construction (see Sec. *CPU Data Structure*), the tree traversal may have to stop at a leaf node before reaching the desired level. However, since the root node is always present, line 5 of Algorithm 1 will always return a node containing the sampling position.

Once the appropriate node has been retrieved, the intensity value at the sampling position has to be determined for each channel. If the node is flagged as homogeneous, we use the stored AVG value of the current channel. If the node is not homogeneous, i.e., it is associated to a brick, we have to distinguish whether the corresponding brick is present on the GPU. In case of an available brick the intensity value is determined by trilinear filtering and the brick is flagged as used, whereas a missing brick is requested and approximated by its AVG value. The intensity value at the sampling position is mapped via a transfer function to a color value and composited depending on the rendering mode. Schubert et al. [24] have summarized and compared different approaches for rendering multi-channel volumes. According to their classification we use an accumulation level intermixing approach, which provides the best visual results.

Before the next rendering pass requested bricks are uploaded onto the GPU depending on the available brick buffer slots. A problem of this basic approach is the partially bad image quality depending on the order of brick requests. Bricks requested first are uploaded into the brick buffer resulting in rendered areas of full quality and areas that are completely approximated by AVG values since the requested bricks could not be uploaded into the full brick buffer. This problem is solved in two different ways, depending on the user interaction as described in the following two sections.

*Full-Frame Ray-Casting*
During user interaction and the first rendering passes of an image we are using the full-frame ray-casting to achieve a uniform image quality with interactive frame rates. The basic approach is modified by no longer approximating missing non-homogeneous bricks by their AVG values. Instead the algorithm checks whether a brick of one of the two previous traversed tree levels is present in the brick buffer. If this is the case, this brick is used as approximation. If not, a fallback to the AVG values of the node is used. Nevertheless, all three bricks of the current and previous two tree levels are requested. During the brick upload bricks of higher LODs are preferred. Thus, the entire data set can be visualized at the best LOD which fits into the brick buffer. No AVG values are used as long as at least $\frac{1}{64}$th of the data set size fits into the brick buffer.

*Refinement Ray-Casting*
If no user interaction takes place and a first image has been rendered by the full-frame ray-casting at a potentially lower LOD, we start the refinement ray-casting to obtain an image at full quality. Leaving the old image persistent on the screen, we refine the entire screen or just parts of it. This time, missing bricks are not approximated by the AVG values or bricks of lower LODs. Instead, after requesting the missing brick, the current ray position and ray color are cached in a refinement buffer. In the next rendering pass the new rays continue from the last cached positions. If all rays are successfully terminated, i.e., no bricks are requested during the last rendering pass, the refinement rendering will replace the old rendered image. Thus, the image quality is ever-improving as long as the camera position is not changed. However, since this approach does not support interactive frame rates, it is only used after a first approximation has been obtained by the full-frame rendering.

*Filtering across Brick Boundaries*
All volume rendering techniques based on octree structures and hierarchical approaches face the problem of trilinear filtering across brick or hierarchy level boundaries. Although trilinear filtering can be performed within a brick, visual fragments can occur on the border between two bricks, since not all voxels that are needed for the filtering are present in the currently sampled brick. The trivial solution would be to determine the missing border brick during ray-casting and also request this brick. However, this approach is not suitable since determining the associated border brick results in expansive octree traversals and more brick uploads. Overall, the performance impact would be too negative. The solution we are using is to increase the brick resolution and overlap two bricks by one voxel as used by Engel [18] or Gobetti et al. [17]. Thus, no more bricks have to be requested and the ray-casting performance is not influenced. However, the needed memory is increased (by 20% for $32^3$ bricks) through the redundant storage of border voxels in two bricks. The needed memory could be reduced by using only one border voxel as shown by Weiler et al.[25], but would result in potentially more requested bricks during ray-casting. However, these approaches increase the incremental construction time, since neighboring bricks have to be loaded for each border update at each octree level.

Thus, we calculate the brick borders in a separate background thread after the octree construction has finished. Our tests have shown that rendering artifacts caused by the missing trilinear filtering during octree construction are not overly disturbing and do not hamper the scientists in their initial inspection of the data set. However, if the octree is stable and the background thread has finished, we provide a correct trilinear filtering.

The problem of filtering across hierarchy levels has been addressed for instance by Beyer et al. [26]. At this point, we have not implemented one of these approaches, since we have not noticed perceivable rendering artifacts in our test cases.

*Correction of Chromatic Aberration*
Data sets created by light microscopy or even electron microscopy suffer from chromatic aberration [27]. Channels of different light wavelengths are reflected in a slightly different way and result in shifts between channels. If this shift is not properly corrected by the microscope itself, the data set must be corrected by the rendering software. A straightforward solution would be to re-sample each channel by user defined parameters, e.g., as done by Imaris [3]. This approach is impractical for us, since it results in long preprocessing times as the channel re-sampling takes time. If the user-defined parameters are set incorrectly, the re-sampling has to start over. Also the re-sampling requires the presence of the entire data set, which is not the case during incremental construction.

Thus, we have decided to perform the correction during the rendering. We store for each channel a separate transformation matrix. The matrix can be used to model any linear transformation, i.e, translation, scaling, and rotation. However, in our use cases the most common corrections are translations and small scalings. Thus, each sampling point on the ray is multiplied by the transformation matrix associated with the current channel to get the correct data point in the data set. This technique allows the correction of chromatic aberration on-the-fly without preprocessing and without the presence of the entire data set. Also in a "what you see is what you get" manner, the biologists can try to find the most suitable correction parameters by shifting channel textures in the 2D slice views.

### Data Transfer
The octree construction and ray-casting require the transfer of different types of data between the storage levels as illustrated in Fig. 5. During the incremental construction new allocated bricks have to be stored on disk and changes in the octree structure have to be transfered to the GPU. During the ray-casting bricks have to be transfered from disk to GPU memory using the RAM as cache. These transfers are conducted before each rendering pass.

*Brick Upload from RAM to GPU*
For this subsection we assume a completely constructed octree and the entire brick pool to be present in the RAM. Since brick buffer slots on the GPU are limited, we have to replace previously uploaded bricks by newly requested ones. To determine the bricks to be replaced, the CPU downloads the flag buffer from the GPU and evaluates it. In the refinement mode we can replace all buffer slots, since the bricks are no longer needed for the current refinement rendering pass. The full-frame mode, on the other hand, requires a more sophisticated brick upload strategy, since bricks that have been used during the last rendering pass are most likely also required in the subsequent one. Therefore, we try to replace only unused bricks by requested ones. However, if the number of requested bricks exceeds the number of unused brick buffer slots, requested bricks at a higher LOD, which cover a larger volume space, are prioritized to replace lower level bricks, even if these were used during the last rendering pass. Interactivity is achieved in both rendering modes by limiting the amount of time per frame used for data uploads to the GPU. Empirical tests have shown a time of 100 ms to 200 ms as best trade-off between interactivity and rendering quality improvement per frame. We have to point out that each brick buffer update requires a node buffer update, since each upload or removal of a brick has to be reflected by an update of its associated node entry.

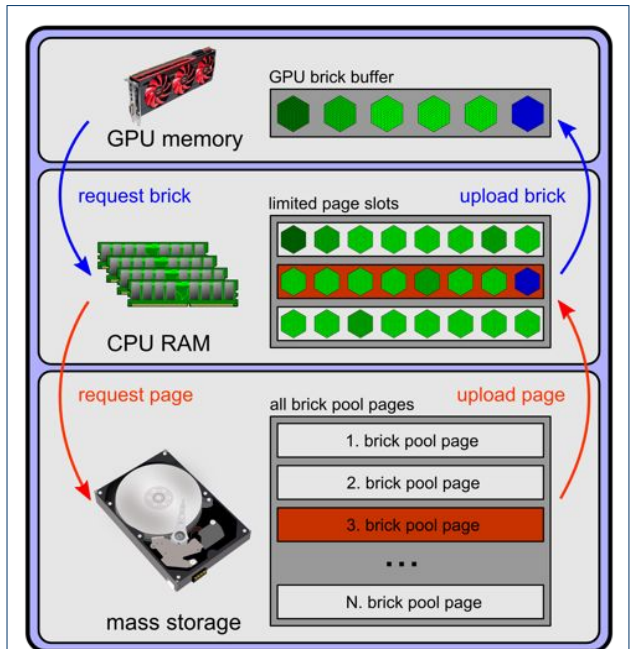

**Figure 5 Illustration of the interaction between the memory hierarchy levels.**
Triggered by the rendering algorithm on the GPU a missing brick is requested from the CPU. If the brick is not present in the RAM yet, the request is forwarded to the disk provoking an upload of the brick pool page containing the requested brick to the RAM. Only the needed brick of the uploaded page is passed to the GPU.

*Brick Upload from Disk to RAM*
The previous assumption that all bricks should be present in the RAM is in general not sustainable, since the required brick pool memory can exceed the available RAM. Thus, parts of the brick pool have to be uploaded from disk. Since disk accesses are expensive, we have chosen a paging approach illustrated in Fig. 5. Multiple bricks are gathered in one *brick pool page*. If the GPU requests a brick, which is not present in the RAM yet, the entire brick pool page is uploaded as shown in red in Fig. 5. Since only a limited number of pages are kept in the RAM, the least recently used page has to be removed from RAM if this limit is reached. If this page contains at least one modified brick, it has to be saved to disk before removal. If a brick of a page is in use, e. g., by the ray-caster or construction thread, the page can not be removed from RAM. Thus the construction and rendering threads have to handle unresolved brick requests if all brick buffer pages are in use. An unresolved brick request by the GPU was already described in Sec. *Basic Ray-Casting*. The construction threads fall asleep when waiting to receive the requested brick, since a skipping of the brick is not possible.

*Incremental Octree Updates on the GPU*
The previous two subsections assumed the octree to be completely constructed during rendering. To reflect updates of the octree structure during incremental construction, e.g., newly added nodes or changed AVG values, the initially uploaded node buffer on the GPU has to be adjusted. The straightforward approach would be to construct a new node buffer from the current octree and replace the old buffer on the GPU. Empirical tests have shown that this approach is inappropriate, since the tree traversal to construct the new node buffer costs too much time. Besides, an update of all nodes would cause the deletion of all bricks on the GPU, since all bricks could be out-of-date, and would cause flickering in the rendering, since the rendering would always start with a cleared brick buffer. Therefore, the octree tracks all changes to its structure during incremental construction. These changes are applied to the node buffer before each rendering pass. There are three kinds of change events: node creation, node deletion and node update. A creation event adds a new node entry at the associated position in the node buffer and updates the child pointer of the parent node entry. Since the node buffer has been initialized for a complete octree (see Sec. *GPU Data Structure*), no other node entries have to be moved or modified. A deletion event removes the node entry from the buffer and clears the child pointer of the parent node entry. A node update event updates the associated node entry's child group pointer and AVG values. Since a brick already loaded to the GPU would be out of date, the *brickInBrickBuffer* flag is set to false. Thus, a re-upload of modified bricks is enforced.

Since only modified node entries are transmitted to the GPU, unmodified nodes and bricks can remain untouched on the GPU minimizing potential flickering.

**Table 1** Data sets and results of the performance benchmarks. The three channels of datasets D2 and D3 are Alexa Fluor 488, Alexa Fluor 568 and Alexa Fluor 647. All files are in the OME-TIFF format.

| Name | Dimensions | Channels | Data Size | Brick Pool Size | Construction Time (min) | Full-Frame DVR / MIP (FPS) | Refinement DVR / MIP (sec) |
|---|---|---|---|---|---|---|---|
| D1 (RAM) | $2560 \times 2160 \times 751$ | 1 | 7.9 GB | 3.6 GB | 2 : 56 | 23 / 27 | 3.6 / 5.1 |
| D1 (DISK) | | | | | — | — | — |
| D2 (RAM) | $1004 \times 1002 \times 1611$ | 3 | 10.3 GB | 9.1 GB | 2 : 05 | 11 / 15 | 9.1 / 9.9 |
| D2 (DISK) | | | | | 2 : 15 | 11 / 15 | 13.1 / 14.2 |
| D3 (RAM) | $1004 \times 1002 \times 3315$ | 3 | 23.8 GB | 18.6 GB | — | — | — |
| D3 (DISK) | | | | | 5 : 59 | 7 / 12 | 214.2 / 76.5 |

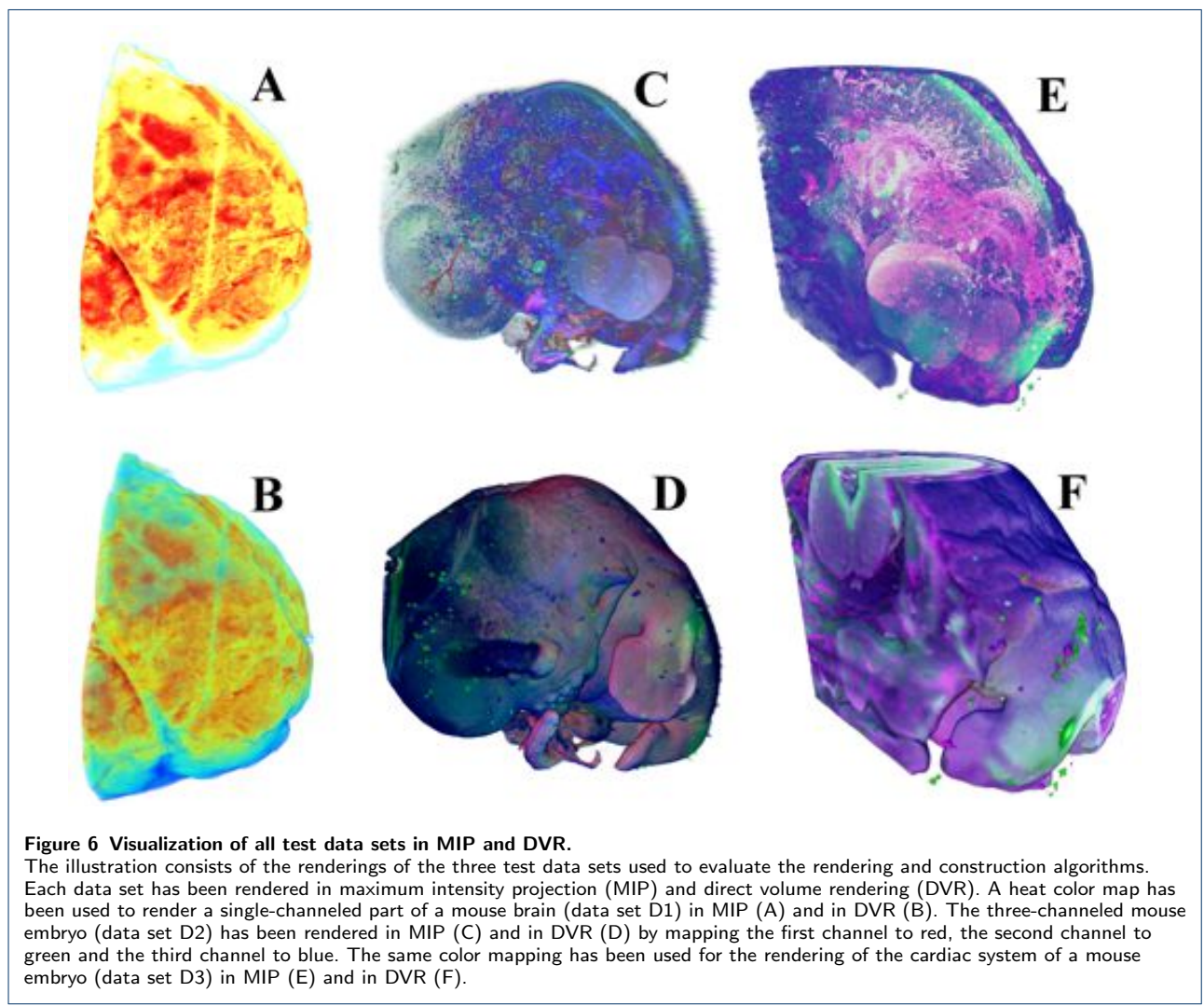

**Figure 6 Visualization of all test data sets in MIP and DVR.**
The illustration consists of the renderings of the three test data sets used to evaluate the rendering and construction algorithms. Each data set has been rendered in maximum intensity projection (MIP) and direct volume rendering (DVR). A heat color map has been used to render a single-channeled part of a mouse brain (data set D1) in MIP (A) and in DVR (B). The three-channeled mouse embryo (data set D2) has been rendered in MIP (C) and in DVR (D) by mapping the first channel to red, the second channel to green and the third channel to blue. The same color mapping has been used for the rendering of the cardiac system of a mouse embryo (data set D3) in MIP (E) and in DVR (F).

## Results and Discussion

All performance tests have been done on a PC with an Intel Core i7-2600K CPU @ 3.40GHz processor, AMD Radeon 7900 Series graphic board and Samsung SSD 840 PRO disk. The meta information of all data sets and performance benchmark results are listed in Table 1. A visualization of all data sets can be seen in Fig. 6. During construction of the octrees and rendering either the data sets were stored entirely in the RAM (RAM), or the available RAM for their storage was limited to 4 GB in order to enforce disk accesses (DISK). The full-frame FPS has been taken from rotating the data set by 360 degrees and averaging the rendering time over 100 rendering passes. The refinement time determines the time until the refinement is complete. For all tests we used a 512 MB brick buffer size on the GPU, a brick resolution of $64^3$, a homogeneity threshold of 5% of the data set intensity range and a screen resolution of $1024 \times 1024$.

For all three data sets the MIP full-frame FPS are significantly higher than the DVR results. This is caused by the number of transfer function texture look-ups. A texture look-up has to be performed only once in MIP, whereas DVR needs a texture access in every sampling step. The forced disk accesses have no significant impact on the full-frame rendering FPS, since at some point, a stable brick configuration of higher level bricks are on the GPU and no new bricks have to be uploaded during the rotation. However, the disk streaming has a higher performance impact in the refinement mode, since nearly the entire brick pool is uploaded to the GPU in this mode, which results in a large number of page switches. The construction time is only slightly affected, since each page is usually just loaded once into the RAM. For the refinement itself, DVR was faster than MIP, since early ray termination can be used. Only for data set D3 the refinement rendering was slower. A possible explanation is that too many page swaps have occurred.

**Data set D1:** *Part of a mouse brain (Fig. 6 A+B).* The sparse nature of this data set results in an optimized octree with many homogeneous bricks. Often no bricks have to be requested by the rendering algorithm and the AVG values can be used. Thus, the data set yields nearly realtime rendering performance with over 20 FPS. Since the optimized brick pool size is below 4 GB the DISK case would be redundant and can be skipped.

**Data set D2:** *Entire mouse embryo (Fig. 6 C+D).* Since this data set has three channels, the number of homogeneous bricks is reduced in comparison to D1 and the optimized brick pool size is still 9.1 GB. This results in lower FPS and longer refinement times. We would like to point out the shorter construction time than for D1, although the data set size is larger. This is explained by our thread implementation which starts a construction thread for each channel. Thus, the single channel data set does not benefit from the potential thread usage. The alternative to start a thread for each octree octant has been tested and results in worse construction times than a single threaded construction. Each thread requests different bricks, resulting in more disk accesses, and cyclic buffer reloads can occur. Also threads have to wait to receive requested bricks, since although they are started in different octants, they will update the same channel of the same nodes at a certain tree level (e. g., the root).

**Data set D3:** *Cardiac system of a mouse embryo (Fig. 6 E+F).* With an optimized brick pool size of 18.6 GB this data set was too large to be entirely stored in the RAM. However, the full-frame rendering is still interactive.

## Conclusion

We have introduced our technique for rendering large volumetric multi-channel microscopy data streams at interactive frame rates on commodity hardware. By using our novel incremental octree construction technique the implemented system enables the user to interact with the progressively rendered volume during data acquisition and to interrupt flawed scans early on. The initial tests with our cooperation partners resulted in very positive feedback, and it is planned to replace their commercial software in the near future.

As a next step of the development we are trying to improve the OpenCL kernels by optimizing the work group memory usage to perform at higher FPS and to handle larger data sets. Also, we will implement the trilinear filtering across different hierarchy levels as described in Sec. *Filtering across Brick Boundaries*.

It is worth mentioning that the needed memory could be reduced by storing only the non-homogeneous channel data of each brick instead of storing always all channels, whether single channels are homogeneous or not. This approach would result in not equally sized bricks which have to be handled by the rendering and storing algorithms. Especially the 64 bits of node entries (see Section *GPU Data Structure*) would have to be increased, since AVG values and brick pointers must be stored simultaneously. Also new flags would have to specify which channels are present in the brick and which are represented by their AVG values. The increased node entries would result in more memory transfers during octree traversal on the GPU. For future work we will benchmark the impact of decreased brick sizes related to increased node entries and more administration effort.

We will also have a look at compression methods and try to combine them with our octree approach. Although these methods have not been considered in our design yet, they could be a way to reduce the memory transfer between the different storage levels. Most promising are the local wavelet compression by Nguyen et al. [28] or the tensor approximation by Kolda et al. [29].

**List of abbreviations used**
AVG: *average (value of a brick)*
CPU: *central processing unit*
DVR: *direct volume rendering*
FPS: *frames per second*
GPU: *graphics processing unit*
LOD: *level of detail*
MIP: *maximum intensity projection*
PC : *personal computer*
RAM: *random access memory*

**Competing interests**
The authors declare that they have no competing interests.

**Acknowledgements**
We thank Friedemann Kiefer and his work group from the Max Planck Institute for Molecular Biomedicine in Münster for sharing their expertise in cellular biology and for providing the used data sets. This work was partly supported by the Deutsche Forschungsgemeinschaft, CRC 656 "Cardiovascular Molecular Imaging" (projects Z1 and Ö).